\documentclass[letterpaper,twocolumn,10pt]{article}
\usepackage{usenix}
\usepackage{titling}
\usepackage{tikz}
\usepackage{multirow} 
\usepackage{pifont}
\usepackage{amsmath}
\usepackage{amssymb}
\usepackage[normalem]{ulem}
\usepackage{url}

\usepackage{subcaption}
\usetikzlibrary{tikzmark}
\usepackage[most]{tcolorbox}
\usetikzlibrary{calc,fit,backgrounds}

\newtcolorbox{algobox}[1]{
    colback=gray!5, colframe=gray!50, 
    boxrule=0.5pt, arc=2pt,
    title={\small #1}, fonttitle=\bfseries
}

\usepackage{graphicx}
\usepackage{booktabs}
\usepackage{algpseudocode}
\usepackage[linesnumbered,ruled,vlined]{algorithm2e}
\usepackage{tabularx}
\usepackage{authblk}
\usepackage{tablefootnote}
\usepackage{float}
\usepackage{xspace}
\usepackage[linesnumbered,ruled,vlined]{algorithm2e}
\usepackage[utf8]{inputenc}
\usepackage{pythonhighlight}

\usepackage{enumitem} 
\usepackage{pifont}
\usepackage{placeins}
\usepackage{comment}
\usepackage{cleveref}
\usepackage[frozencache,cachedir=minted-cache]{minted}
\usepackage{booktabs}
\usepackage{makecell}

\newcommand{\mixed}{\ding{51}\kern-0.62em\ding{55}} 

\usepackage{tikz}

\newcommand{\sys}{BatchGen\xspace}

 \usepackage{etoolbox}
 
 \makeatletter
 \patchcmd{\maketitle}
 	{\@maketitle}
 	{\vspace{-8em}\@maketitle\vspace{-6em}}
 	{}{}
 \makeatother

 \makeatletter
 \patchcmd{\@startsection}
  {#4}
  {#4\addtolength{\@tempskipa}{-1pt}}
  {}{}
 \makeatother

\algdef{SE}[LOOP]{Loop}{EndLoop}[1]{\textbf{loop} #1}{}

\begin{document}
\pagestyle{empty}


\newcommand{\para}[1]{\noindent\textbf{#1}}

\newcommand{\tinyskip}{\vspace{3pt}}
\newcommand{\mypar}[1]{\tinyskip\noindent\textbf{#1.}\xspace}
\newcommand{\myitem}[1]{\item\textbf{#1.}\xspace}

\newcommand*\myc[1]{%
\scalebox{0.78}{\begin{tikzpicture}[baseline=-3pt]
  \node[draw,circle,inner sep=0.5pt, fill=black] {\textcolor{white}{\textsf{\textbf{#1}}}};
\end{tikzpicture}}}


\makeatletter
\DeclareRobustCommand\onedot{\futurelet\@let@token\@onedot}
\def\@onedot{\ifx\@let@token.\else.\null\fi\xspace}

\newcommand{\aff}[1]{\textsuperscript{#1}}

\definecolor{equalgreen}{HTML}{00CC00}

\def\eg{\textit{e.g}\onedot} \def\Eg{\textit{E.g}\onedot}
\def\ie{\textit{i.e}\onedot} \def\Ie{\textit{I.e}\onedot}
\def\cf{\textit{c.f}\onedot} \def\Cf{\textit{C.f}\onedot}
\def\etc{\textit{etc}\onedot} \def\vs{\textit{vs}\onedot}
\def\wrt{w.r.t\onedot} \def\dof{d.o.f\onedot}
\def\etal{\textit{et al}\onedot}
\makeatother
\newcommand{\Sum}[3]{\sum\limits_{#1}^{#2}{#3}}
\newcommand{\lr}[3]{\left #1 {#3} \right #2}
\newcommand{\E}[2]{E[#1,#2]}

\newcommand{\fixme}[1]{{\color{red}[#1]}}
\ifdefined\RELEASE
  \newcommand{\ignore}[1]{}
  \newcommand{\luo}[1]{}
  \newcommand{\leyang}[1]{}
  \newcommand{\zhan}[1]{}
  \newcommand{\TODO}[1]{}
  \renewcommand{\thefootnote}{\arabic{footnote}}
\else
  \newcommand{\ignore}[1]{}
  \newcommand{\review}[1]{{{#1}}}
  \newcommand{\luo}[1]{{\textcolor{gray}{[~LUO:~#1~]}}}
  \newcommand{\leyang}[1]{{\noindent\textcolor{blue}{[LX:~#1]}}}
  \newcommand{\tairan}[1]{{\noindent\textcolor{blue}{[Tairan:~#1]}}}
  \newcommand{\zhan}[1]{{\textcolor{cyan}{[~Zhan:~#1~]}}}
  \newcommand{\lx}[1]{{\noindent\textcolor{teal}{[Le:~#1]}}}
  \newcommand{\jysc}[1]{{\noindent\textcolor{red}{[jysc:~#1]}}}
  \newcommand{\TODO}[1]{{\textcolor{red}{TODO:~#1}}}
  \renewcommand{\thefootnote}{\arabic{footnote}}
\fi

\newenvironment{tightlist}{
\begin{list}{$\bullet$}{
    \setlength{\topsep}{.1em}
    \setlength{\partopsep}{0in}
    \setlength{\parskip}{0in}
    \setlength{\itemsep}{0in}
    \setlength{\parsep}{0in}
    \setlength{\leftmargin}{1em}
    \setlength{\rightmargin}{0in}
    \setlength{\itemindent}{0in}
}}
{\end{list}}

\newenvironment{tightenum}{
\begin{list}{\myc{\arabic{enumi}}}{
    \usecounter{enumi}
    \setlength{\topsep}{.1em}
    \setlength{\partopsep}{0in}
    \setlength{\parskip}{0in}
    \setlength{\itemsep}{0in}
    \setlength{\parsep}{0in}
    \setlength{\leftmargin}{1.5em}
    \setlength{\rightmargin}{0in}
    \setlength{\itemindent}{0in}
}}
{\end{list}}

\title{\Large \bf \sys: An Architecture for Scalable and Efficient Batch Inference}
\date{}

\author{
Tairan Xu$^{\dagger\textcolor{equalgreen}{*}}$,
Leyang Xue$^{\dagger\textcolor{equalgreen}{*}}$,
Zhan Lu$^{\dagger\textcolor{equalgreen}{*}}$,
Jinfu Deng$^\ddagger$,
Hongyang Xiao$^\ddagger$,
Yinsicheng Jiang$^{\dagger}$,
Congjie He$^{\dagger}$,
Matej Sandor$^{\dagger}$,
Le Xu$^{\dagger}$,
Luo Mai$^{\dagger}$
\\
\it $^\dagger$University of Edinburgh, \it $^\ddagger$Tencent
} 

\maketitle
\thispagestyle{empty}

\def\thefootnote{*}\footnotetext{Co-leading authors.}
\def\thefootnote{\arabic{footnote}}

\begin{abstract}
Batch inference has become a central mode of AI computation, yet existing inference engines still rely on execution models designed for interactive serving. When scaled to millions of sequences, batch workloads reveal two fundamental requirements: the ability to handle extreme inter- and intra-sequence load variation that emerges only at runtime, and the ability to sustain high utilization across large fleets of GPUs. Existing systems fail to meet these requirements, losing substantial fractions of achievable throughput.

We introduce a new architectural foundation for batch inference: the sequence coroutine compute model, which represents each sequence as a fine-grained, event-driven coroutine. This model exposes expressive primitives that allow the runtime to reorganize work dynamically, enabling larger expert-level batches, mitigating stragglers, reallocating work across devices, and maintaining utilization even on cost-effective or memory-constrained GPUs.
Building on this abstraction, we implement \sys, a production-ready system that uses the coroutine model at cluster scale. On a 128-GPU cluster, \sys reduces batch completion time by up to $2.3\times$, and on memory-constrained accelerators it outperforms the strongest offloading baseline by up to $9.6\times$. We will open-source \sys at \url{https://github.com/batchgen-project/batchgen}.

\end{abstract}

\begin{tcolorbox}[colback=yellow!10, colframe=black, boxrule=0.5pt,
		arc=2pt, left=4pt, right=4pt, top=4pt, bottom=4pt, enhanced,
		sharp corners, width=\columnwidth, boxsep=2pt]
	\small
	\textbf{Note:} This paper appears in USENIX OSDI 2026.
\end{tcolorbox}

\section{Introduction}

Batch inference has become one of the fastest-growing and most resource-intensive modes of AI computation~\cite{batch-trend,cloud-batch-growth}. Offline inference pipelines~\cite{zhang2025data, naeem2024retclean}, synthetic data generation~\cite{long2024llms, sahu2022data}, model evaluation~\cite{liangholistic, srivastava2023beyond, an2024eval}, test-time scaling~\cite{cot,wangself}, and RL rollouts~\cite{rlhf-rollout} now dominate the compute budgets of major AI deployments. Unlike interactive serving~\cite{vllm,sglang,trt-llm}, these workloads optimize batch completion time (BCT), operate at massive sequence scales~\cite{openaibatch,awsbatch}, and often run on memory-limited accelerators.

Batch inference introduces fundamentally new system requirements. A batch engine must (1) dynamically increase expert-level batch sizes for sparse models, (2) continuously rebalance load as long-tail sequences emerge, and (3) sustain high utilization across large fleets of GPUs. Crucially, these requirements arise directly from how modern AI is being scaled. State-of-the-art models rely on extreme sparsity, especially large Mixture-of-Experts (MoE) architectures, where tokens activate only a few experts among hundreds. Even million-token global batches decompose into small per-expert batches that fall far below GPU-saturating throughput. Meanwhile, test-time scaling and heavy reasoning workloads generate persistent long-tail sequences, creating stragglers that dominate BCT and leave many GPUs idle. These structural properties -- expert sparsity, modular computation, and heavy-tailed generation -- define a new operating mode that demands a scheduler capable of reorganizing computation at fine granularity.

Existing systems cannot meet these requirements because they inherit a latency-first execution model from interactive serving. Systems such as vLLM~\cite{vllm}, SGLang~\cite{sglang}, and TensorRT-LLM~\cite{trt-llm} statically bind each sequence's computation and KV state to a fixed GPU, executing forward passes atomically to minimize per-sequence latency. This prevents pausing between neural modules, forming larger expert batches at runtime, or redistributing load as long-tail sequences appear. Disaggregated designs~\cite{distserve,megascale-infer} retain fixed placement assumptions and thus continue to underutilize GPUs. Even throughput-oriented systems~\cite{nanoflow,moe-lightning} inherit the same static scheduling model and cannot address the root causes of insufficient expert batching or decoding stragglers.

Our key intuition is that sequences themselves provide the right granularity for dynamic scheduling. Neural networks expose natural yield points at module boundaries, and per-sequence state is compact and migratable. This motivates a shift in design: treat sequences not as long-lived tasks pinned to devices, but as fine-grained, event-driven coroutines that can pause, resume, combine, partition, and migrate across GPUs. This architecture directly addresses the structural inefficiencies of batch inference. Specifically, this paper makes the following contributions:

\mypar{(1) Sequence coroutine compute model} Each sequence is represented as a coroutine that carries all state required for correct execution. The model provides four expressive primitives, \texttt{yield}, \texttt{combine}, \texttt{partition}, and \texttt{migrate}. These operations allow the system to reorganize computation while preserving correctness. They make it possible to pause execution and combine sequences to form larger batches for sparse modules, partition stragglers across idle GPUs, and migrate coroutine state for effective load balancing. Existing inference architectures cannot naturally express these behaviors.

\mypar{(2) Efficient sequence coroutine runtime}
The runtime implements event-driven scheduling at cluster scale and is grounded in two key observations.
Prefill and decode place very different demands on memory and therefore benefit from different strategies for managing coroutine state.
Meanwhile, both phases can run within a unified execution framework that jointly schedules, combines, and partitions coroutines over a shared pool of GPUs.
This approach maximizes utilization and avoids leaving devices idle when workloads fluctuate.

\mypar{(3) \sys, a production-ready system that utilizes this design} \sys supports multi-node and multi-GPU execution, provides a compatible and expressive batch inference interface, and includes a checkpoint management service that enables rapid cold starts.
These capabilities are essential for practical deployment, especially when resource availability changes over time.

\mypar{(4) Extensive evaluation of the system} Across MoE models, test-time scaling workloads, and reinforcement learning rollouts, BatchGen delivers strong and consistent improvements. On a 128-GPU cluster, it increases batching efficiency for expert layers, reduces the impact of long-tail decoding stragglers~\cite{rollpacker}, and lowers batch completion time by up to $2.3\times$. It outperforms leading inference engines~\cite{vllm, sglang, trt-llm} even on memory-constrained accelerators.

\sys has been deployed on production clusters to support batch-inference applications. We believe the event-driven sequence coroutine architecture represents a fundamental shift for batch inference, akin to the transition from thread-per-request servers (e.g., Apache~\cite{apache}) to event-driven architectures (e.g., NGINX~\cite{nginx}) that redefined large-scale web systems.

\section{Background and Motivation}

\begin{figure}[t]
    \centering
    \includegraphics[width=\linewidth]{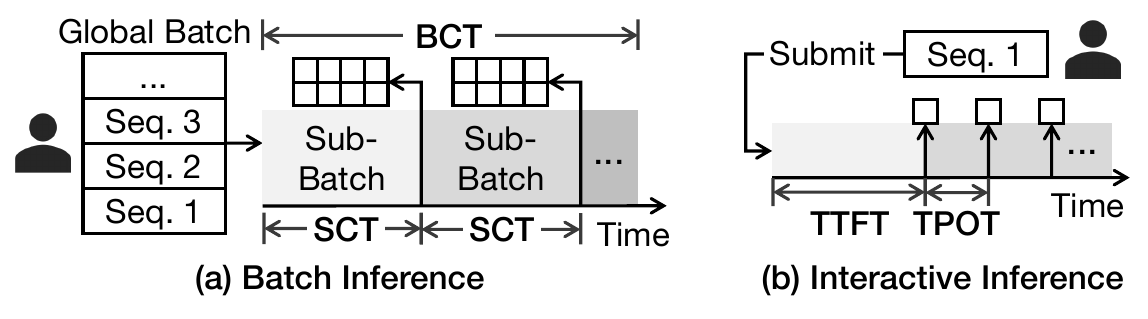}
    \caption{New system design goals for batch inference.}
    \label{fig:inference-mode}
\end{figure}

\subsection{Batch Inference and Its Key Objective}

Major vendors now provide dedicated batch inference services, including OpenAI Batch Inference~\cite{openaibatch}, AWS Bedrock batch processing~\cite{awsbatch}, and Azure batch endpoints~\cite{azurebatch}. 
These services follow a simple model: users submit a batch of input sequences, such as prompts for LLMs, and receive the outputs only after the entire batch has finished processing. Depending on the application, batch sizes can vary widely, from tens of sequences in test-time scaling or thousands in reinforcement learning rollouts to millions in large-scale offline inference.

\mypar{Batch inference systems focus on minimizing the completion time of the entire batch, which is fundamentally different from the responsiveness metrics emphasized in interactive inference} As shown in ~\autoref{fig:inference-mode}, we define the \textit{batch completion time} (BCT) as the total time from batch submission until every sequence in the batch has been processed.
When an application requires partial or early results, the system divides the input into sub-batches, and returns outputs incrementally as each sub-batch finishes. In these cases, we define the \textit{sequence completion time}  (SCT) as the time from batch submission until a specific sequence's output becomes available. SCT is determined by the completion time of the sub-batch that contains the sequence.
Our primary optimization goal is to minimize BCT, since the user receives the full result only when the entire batch is done. In contrast, interactive inference prioritizes user-perceived responsiveness, focusing on metrics such as time to first token (TTFT) and time per output token (TPOT).

\begin{figure*}[!t]
    \centering

    \begin{subfigure}[b]{0.265\textwidth}
        \centering
        \includegraphics[width=\linewidth]{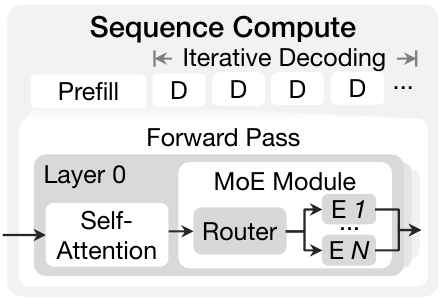}
        \caption{MoE sequence compute.}
        \label{fig:moe}
    \end{subfigure}
    \hfill
    \begin{subfigure}[b]{0.36\textwidth}
        \centering
        \includegraphics[width=\linewidth]{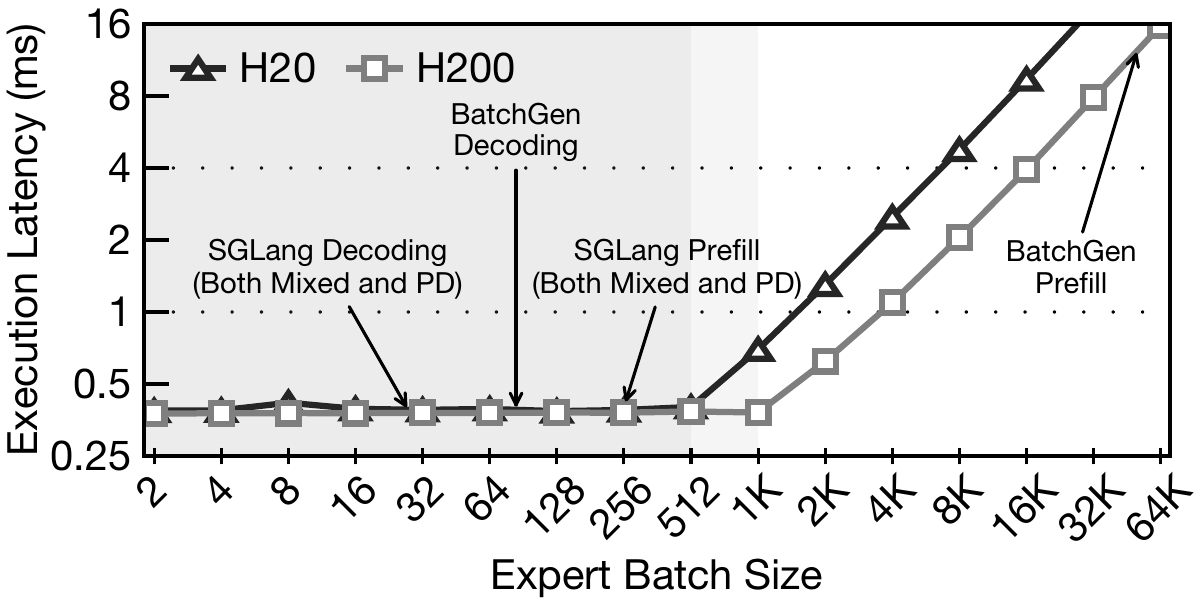}
        \caption{Insufficient batch size for experts.}
        \label{fig:expert-batch}
    \end{subfigure}
    \hfill
    \begin{subfigure}[b]{0.36\textwidth}
        \centering
        \includegraphics[width=\linewidth]{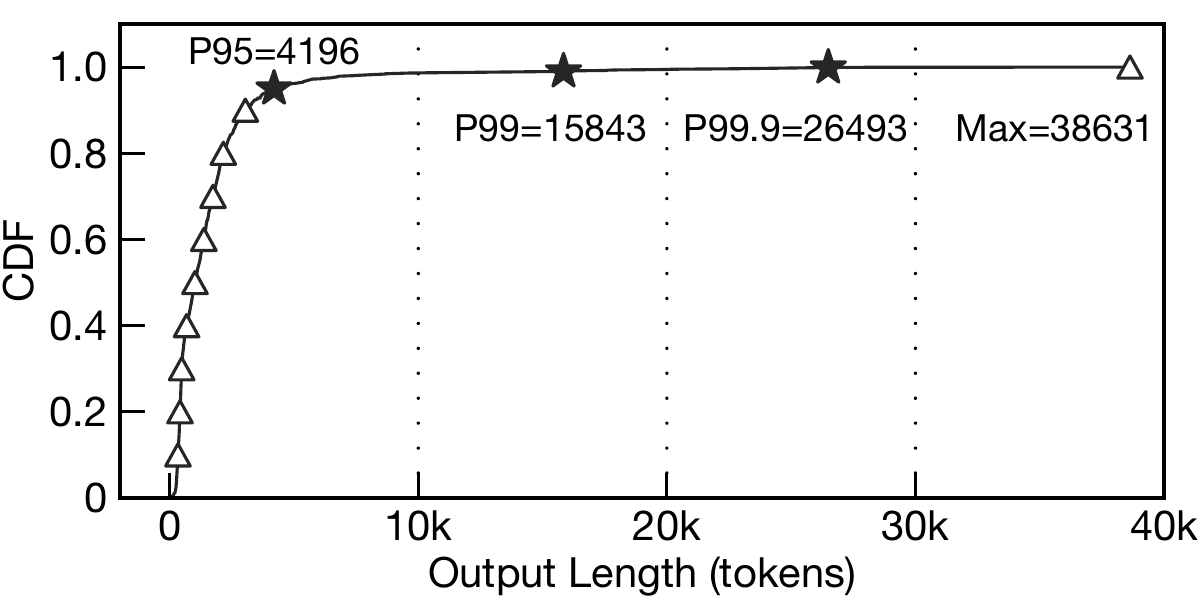}
        \caption{CDF of output lengths for DeepSeek-R1.}
        \label{fig:long-tail}
    \end{subfigure}

    \caption{Challenges in MoE batch inference.}
    \label{fig:three-part-6-7-7}
\end{figure*}

\subsection{The Missing Batch Inference System}
\label{sec:efficiency-scalablity-challenges}

As adoption of batch inference increases across both industry and academic workflows, the demand for an efficient batch inference engine becomes even more evident. However, there are several challenges in scaling batch workloads: 
(i)~batches continue to grow in size, which requires inference engines to sustain very high levels of concurrency.
(ii)~the computation required for each sequence can vary widely, which leads to substantial imbalance across GPUs and directly affects batch completion time. 
(iii)~modern models introduce additional variation even within a single sequence because different neural network modules can have very different computational costs, further increasing the imbalance.

These limitations are not confined to a few specialized workloads. Instead, they appear consistently across batch inference settings. Below, we explain why high concurrency and load imbalance naturally arise from the way modern AI models and applications scale.

\mypar{Insight 1: Intra-sequence imbalance is inherent to sparsity-driven model scaling}
State-of-the-art AI models increasingly rely on sparse architectures, most notably Mixture-of-Experts (MoE), as the primary mechanism for scaling model capacity.
Shown in~\autoref{fig:moe}, modern models such as DeepSeek-R1~\cite{deepseek-r1}, Kimi-K2~\cite{kimi-k2}, GPT-5~\cite{openai2025gpt5systemcard}, Gemini 3 Pro~\cite{pichai2025gemini3}, and Grok~\cite{xai2025grok4modelcard} all deploy MoE models with hundreds of experts. 
The MoE layers contain the majority of the model parameters and contribute a large portion of the compute required for each token.
    
Since each token activates only a small subset of experts, different tokens within the same sequence follow different execution paths. 
Some experts are selected far more often than others, and these variations accumulate as model capacity grows. As a result, intra-sequence imbalance is not a product of specific implementations but an intrinsic consequence of sparsity-driven scaling.

The intra-sequence imbalance causes today's state-of-the-art inference engines, even when they are configured in disaggregated mode (prefill-decoding separation), to suffer from underutilizing the GPUs by around 50\% even though the input batch has sufficiently massive sequences (shown by~\autoref{fig:expert-batch}). 
SGLang mixed means that prefill and decoding phases are deployed on the same servers, while SGLang PD uses prefill-decoding disaggregation, with the two phases deployed on separate servers.

\mypar{Insight 2: Test-time scaling amplifies long-tail generation, causing inter-sequence imbalance}
Batch inference workloads exhibit a persistent long-tail effect, where a small fraction of sequences dominate the overall computation. This phenomenon is especially critical in scenarios where batch completion time dominates the pipeline time, such as RL rollouts and deep-thinking pipelines.

\autoref{fig:long-tail} illustrates this effect for DeepSeek-R1 on a production dataset trace. The imbalance in decoding lengths is severe: the P99 output length is 3.78$\times$ longer than P95, and the maximum length reaches 9.2$\times$ that of P95. These stragglers determine batch completion time, leaving most GPUs idle while waiting for a few long-running sequences.

The long tail issues create two key scalability challenges. (1) First, because decoding is sequential, a few extremely long sequences determine the batch completion time, acting as stragglers. (2) Second, GPUs processing shorter sequences finish early and remain idle, amplifying load imbalance between GPUs.  In production workloads we consistently observe that this imbalance causes existing engines \cite{vllm, sglang, trt-llm}, and their disaggregated variants to lose roughly 10\% to 70\% of achievable GPU performance. This demonstrates that inter-sequence imbalance is not a rare occurrence but an inherent property of large batch inference workloads.

\mypar{Takeaway} Together, these requirements and insights reveal the need for a new class of systems: \textbf{batch-native inference engines} that can sustain high utilization across large fleets of parallel GPUs while processing massive numbers of concurrent sequences, despite inter- and intra-sequence load variation that is only revealed at runtime.

\section{\sys Design Intuitions}

We begin by examining the design philosophy underlying existing inference engines.
\mypar{Today's latency-driven scheduling model is a fundamental mismatch for batch inference}
State-of-the-art systems such as SGLang \cite{sglang} and vLLM \cite{vllm} -- regardless of whether they operate in continuous batching mode, chunked prefill, or a disaggregated configuration -- are architected around a single objective: \textit{complete each sequence as quickly as possible}. To meet this objective, they \textit{bind} each sequence's computation and state (e.g., KV-cache) to a fixed device and execute it without interruption.

This binding philosophy appears at two levels. \textit{Within} a forward pass, execution is atomic: once a batch enters the model, all sequences traverse every layer together before any scheduling decision can be made. There is no mechanism 
to pause between modules--e.g., to accumulate more sequences after attention before executing MoE. \emph{Across} forward passes, sequences remain attached to the same device; moving a sequence's state elsewhere is treated as an exceptional, expensive operation to avoid. As a result, each sequence is permanently bound to its assigned GPU and cannot be moved or rescheduled. Once a sequence joins a batch, it stays with that batch until its own completion.

This design is well suited for interactive serving, where minimizing per-sequence latency dominates all other concerns. However, in batch inference, the rigidity becomes a core limitation-both in the \textit{scheduling plans} the system can express and in the \textit{granularity} at which it can intervene. The engine cannot pause a sequence mid-layer to form a more efficient expert batch, cannot adapt the plan as imbalance develops, and cannot redistribute work when some sequences finish far earlier than others.

\mypar{Requirements for new batch-native inference scheduling model}
In this paper, we revisit batch inference from first principles. In existing systems, a sequence is assigned to specific devices, and this assignment remains fixed because re-binding is viewed as adding unnecessary latency. 
Even in continuous batching~\cite{orca,vllm,sglang}, where micro-batches may be reorganized, the \textit{sequence itself} remains implicitly tied to a particular device and cannot freely move or be decomposed.

This observation leads us to a novel system design opportunity: Can we model sequence computation as fine-grained coroutines and dynamically schedule them in a throughput-oriented, event-driven architecture?

\begin{figure}
    \centering
    \includegraphics[width=1\linewidth]{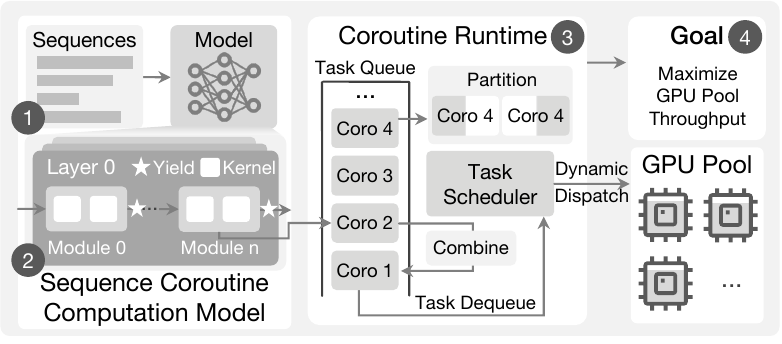}
    \caption{Event-driven Sequence Coroutine Architecture.}
    \label{fig:sequence-coroutine-arch}
\end{figure}

We make four key insights, illustrated in \autoref{fig:sequence-coroutine-arch}, that show why such a system design is practical and effective for today's batch inference workloads:

\begin{tightenum}
    \item \textbf{Sequences provide a natural granularity for coroutine-based scaling.}  
    In batch inference, both compute and memory load scale with the number of sequences, and each sequence has an independent execution lifetime. This makes sequences a natural unit for coroutine computation.

    \item \textbf{Neural network modules offer natural yield points.}  
    Sequence computation performs forward passes through modular components (e.g., attention, MoE layers) in prefill and decode. Module boundaries naturally serve as coroutine yield points for event-driven scheduling.

    \item \textbf{Sequence state can be dynamically combined, partitioned, and migrated.}  
    The state associated with each sequence (e.g., hidden states in prefill, KV cache in decode) is tensor-structured and self-contained, allowing flexible pausing, combining, partitioning, and migration without affecting correctness.

    \item \textbf{Batch inference optimizes batch completion time, hiding the dynamic scheduling latency of sequences.}  
    The extra overhead of coroutine scheduling may be undesirable for interactive workloads, but in batch settings, where end-to-end batch completion time is the primary metric, this overhead is amortized and outweighed by improved throughput from high-concurrency coroutine execution.
\end{tightenum}

To the best of our knowledge, no existing system follows our proposed architectural model. Interactive inference engines such as SGLang~\cite{sglang} and vLLM~\cite{vllm} bind each sequence to a fixed GPU for its entire lifetime. Disaggregated inference systems, including DistServe~\cite{distserve}, Splitwise~\cite{splitwise}, and Mooncake~\cite{mooncake}, relax this constraint by assigning each inference phase, such as prefill or decode, to a specific GPU, but the phases still remain statically placed. MegaScale-Infer~\cite{megascale-infer} further disaggregates decode into attention and expert modules, but continues to rely on fixed placement of these components. Kernel oriented systems such as NanoFlow~\cite{nanoflow} improve execution efficiency, but do not address their underlying scheduling and placement limitations.

\begin{table}[t]
  \centering
  \resizebox{\columnwidth}{!}{
  \setlength{\tabcolsep}{4pt}
  \renewcommand{\arraystretch}{1.1}
  \begin{tabular}{lccc}
    \toprule
    & \makecell{\textbf{Interactive}\\\textbf{Inference}} 
    & \makecell{\textbf{Disaggregated}\\\textbf{Inference}} 
    & \makecell{\textbf{Sequence Coroutine}\\\textbf{(Ours)}} \\
    \midrule
    \textbf{Scheduling plan} &
      Static &
      Static &
      \textbf{Adaptive} \\
    \textbf{Scheduling granularity} &
      Coarse-grained, fixed &
      Coarse-grained, fixed &
      \textbf{Fine-grained, flexible} \\
    \textbf{Operational cost} &
      Medium &
      High &
      \textbf{Low} \\
    \textbf{Throughput} &
      Medium &
      Medium &
      \textbf{High} \\
    \bottomrule
  \end{tabular}
  }
  \caption{Scheduling characteristics of three inference system architectures.}
  \label{tab:sched-compare}
\end{table}

\mypar{Architectural advantages} We name our new system architecture the \textit{event-driven sequence coroutine architecture}.
\autoref{tab:sched-compare} summarizes its potential effectiveness. By exploiting sequence-level coroutines and providing system mechanisms for their yield and combination, we can significantly increase batch sizes on sparse operations (\eg each expert in MoE) at runtime and thereby achieve significantly higher GPU utilization than interactive engines (shown in \autoref{fig:expert-batch}).

Meanwhile, when long-tail generation or load imbalance is detected, low-cost event-driven scheduling can yield, migrate, partition, and recombine sequence coroutines to avoid stragglers and maintain balanced loads. 

Because GPUs are managed as a shared resource pool rather than being pre-allocated to specific sequences, sequence coroutines can be dynamically dispatched to any available GPU, making the system easy to deploy across batch jobs with widely varying input--output length distributions.

Finally, sequence coroutines execute under an event-driven model in which GPUs always have useful events to process, while the scheduler keeps most coroutine state in host memory. This substantially reduces peak GPU memory usage and lowers the operational cost of batch inference.

A powerful analogy is the transition from thread-per-connection servers (e.g., Apache HTTP Server~\cite{apache}) to event-driven architectures (e.g., NGINX~\cite{nginx}, SEDA~\cite{seda}). Apache performs well under modest load but degrades under highly variable, large-scale traffic; NGINX treats requests as coroutines and uses event-driven scheduling to sustain scalability. Batch inference demands the same paradigm shift.

\section{The Sequence Coroutine Model}
\label{sec:coroutine-model}
\subsection{The Sequence Coroutine Abstraction}
\label{subsec:seq-coroutine-definition}

We abstract sequence coroutines as follows: a representation of a neural network model's per-sequence execution that can be paused, migrated, combined, partitioned, and resumed without losing correctness. 

Establishing this abstraction requires answering three key questions:
(i)~How to model the sequence's computation process and intermediate state?
(ii)~How to enable sequence coroutine execution in a neural network that is typically built using modular abstractions?
(iii)~How to execute concurrent sequence coroutines on a limited number of compute units?

\begin{figure}[t]
\centering
\begin{subfigure}{\linewidth}
 \centering
    \begin{minted}[
    fontsize=\fontsize{8}{9}\selectfont,
    frame=single,autogobble,breaklines
    ]{python}
class SequenceCompute:
  seq_id: int; max_out: int; max_in: int; tokens: str
  phase: Literal["prefill", "decoding"]
  status: Literal["init", "active", "inactive", "done"]
  state: object; output: object

  async def yield(...): # Scheduler takes control 
  async def combine(...): # Multiple states
  async def partition(...): # Model parallelism
  async def migrate(...): # Physical location
  async def run(...): # Call module forward

  def callback(...): # Custom local state handling
\end{minted}
\caption{Sequence computation states and functions.}
\label{fig:sequence-compute}
\end{subfigure}
\begin{subfigure}{\linewidth}
 \centering
    \begin{minted}[
    fontsize=\fontsize{8}{9}\selectfont,
    frame=single,autogobble,breaklines,
    escapeinside=||
    ]{python}
def module_wrapper(module: |{\color{blue}\bfseries Module}|):
  async def coroutine(
    s_instance: |{\color{blue}\bfseries SequenceCompute}|, 
    combine_with: List[|{\color{blue}\bfseries SequenceCompute}|] = None,
    n_part: int = 1, part_mode: Literal["TP"]):
    # Scheduler implementation
    return s_instance
  return coroutine
# applies to attention, MoE, prefill, decoding
attention.forward = module_wrapper(attention)
\end{minted}
\caption{Module wrapper definition and coroutine creation.}
\label{fig:module-wrapper}
\end{subfigure}
\begin{subfigure}{\linewidth}
    \centering
    \includegraphics[width=\linewidth]{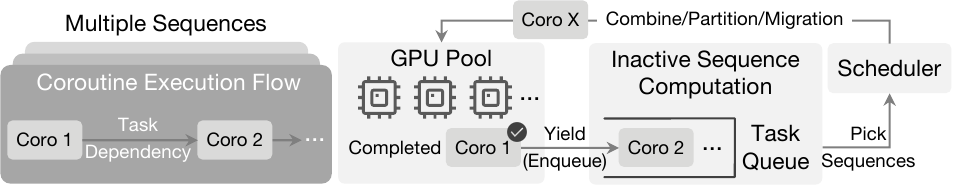}
    \caption{Concurrent sequence coroutine execution.}
    \label{fig:sequence-concurrent}
\end{subfigure}
\caption{Sequence coroutine abstraction.}
\end{figure}

\mypar{Modeling sequence computation and state} 
The sequence computation (defined in \autoref{fig:moe}) requires states that fully determine the current and future computations.
We define the data structure as shown in \autoref{fig:sequence-compute}, using language models as an example for the values in each field.
Each sequence can be uniquely identified, with the amount of computation to be scheduled (\ie \texttt{max\_out}, \texttt{max\_in}).
The coroutine state stores the KV-cache, while the output consists of the hidden states passed between module forward calls, enabling the coroutine to resume execution correctly at the next scheduling point.
Sequence computation \review{exposes} primitives (\ie \texttt{yield}, \texttt{combine}, \texttt{partition}, \texttt{migrate}) to enable coroutine scheduling on sequence computation, and callbacks for customized operations on sequence states.

\mypar{Creating coroutines through module wrappers}
We provide an abstraction layer that preserves the modular semantics of existing model definitions (\eg \texttt{torch.nn.Module}), while inserting scheduling points at runtime (\autoref{fig:module-wrapper}).
During model initialization, \sys applies wrappers on selected modules, allowing coroutine steps to be generated automatically. 
The wrapper allows any custom scheduling policy to be implemented using coroutine primitives.
By default, the coroutine yields the sequence compute on exit.

\mypar{Concurrent sequence coroutine execution}
Concurrency is managed by a coroutine scheduler for coordination over the GPU pool (\autoref{fig:sequence-concurrent}). 
Once modules are wrapped, a sequence computation naturally forms a \emph{coroutine execution flow} that defines the successor relationships between computation stages. When a coroutine yields, it enqueues its successor, while the scheduler selects one or more inactive sequence coroutines from the global queue to dispatch onto available compute units. A many-to-one mapping corresponds to \texttt{combine}, a one-to-many mapping corresponds to \texttt{partition}, and a one-to-one mapping corresponds to a simple resume.

\subsection{Sequence Coroutine Mechanisms}\label{sec:coroutine-mechanisms}
\label{subsec:seq-coroutine-mechanisms}

When designing mechanisms for sequence coroutines, we have three requirements:
(i)~they must provide yield primitive that preserves correctness for stateful neural network execution;
(ii)~they must enable coroutine combination to inflate expert-level batch sizes in MoE computation; and
(iii)~they must support coroutine partitioning to mitigate stragglers in long-tail prefill and decoding stages.

\begin{figure}[t]
    \centering
    \includegraphics[width=\linewidth]{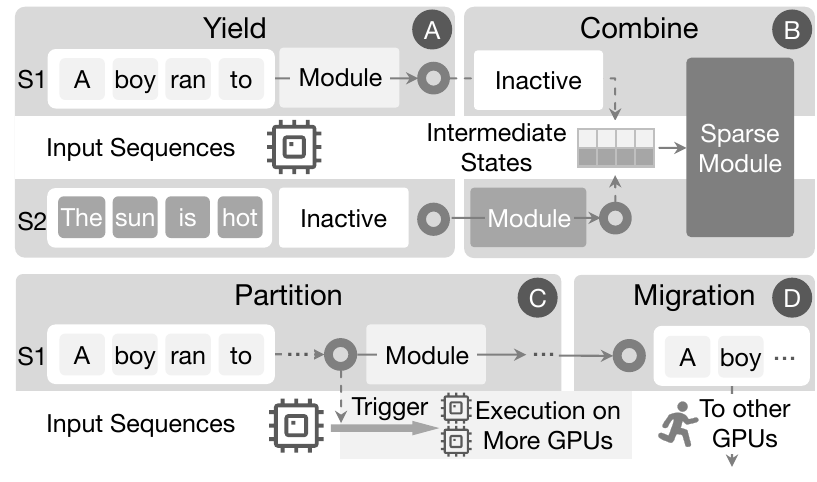}
    \caption{Coroutine primitives used on sequence.}
    \label{fig:coroutine}
\end{figure}

\mypar{Sequence coroutine yield mechanism}
A \texttt{YIELD} in sequence compute behaves analogously to the await primitive in programming languages such as Python: yield suspends a sequence coroutine, checkpointing its state produced and releasing GPUs, such that control can safely return to the scheduler.
Figure~\ref{fig:coroutine}~\myc{A} illustrates this process: 
sequence $S_1$ yields after a module computation and becomes inactive, allowing the scheduler to resume inactive $S_2$ on the same GPU.

\mypar{Sequence coroutine combination mechanism}
Standard coroutine semantics require an explicit \texttt{resume} primitive to continue suspended execution. In our model, resume semantics are part of \texttt{COMBINE}: since the purpose of combining sequences is to form a batch for computation, resumption occurs implicitly. 
This design eliminates a separate primitive while preserving expressiveness.

Analogous to data-stream operators in Spark~\cite{zaharia2012resilient}, \texttt{COMBINE} merges multiple yielded sequences into a single batch for efficient GPU execution. Figure~\ref{fig:coroutine}~\myc{B} illustrates this process: intermediate states (in the form of tensors) are concatenated from $S_1$ and $S_2$, doubling the batch size for the sparse module and increasing compute density on a single GPU.

\mypar{Sequence coroutine partitioning mechanism}
\texttt{PARTITION} distributes a single long-running sequence across multiple GPUs using tensor parallelism to accelerate completion.
Each GPU executes the same program over its assigned portion of the data.
Figure~\ref{fig:coroutine}~\myc{C} illustrates this on a single sequence:
the scheduler finds another idle GPU to form a parallelism group for the module. The GPU group can be local to a node or distributed, depending on the scheduler's decision.

\mypar{Sequence coroutine migration mechanism}
\texttt{MIGRATE} transfers a sequence's state (\eg KV-cache and metadata) to a different physical device for load balancing.
The call is asynchronous while still ensuring consistency.
The physical location of the sequence is left to the scheduler for book-keeping. 
\autoref{fig:coroutine}~\myc{D} illustrates this for moving to another GPU. This also applies to sequences in Host and other media as well.

\subsection{How to Use Sequence Coroutine?}
\label{subsec:optimizing-coroutine-yield-points}

In this subsection, we use an MoE model to illustrate how to use the sequence coroutine abstraction for batch inference.

\mypar{Two types of yield points}
\review{
  \emph{Intra-forward} yield points occur within a single forward pass. 
  When a downstream module reaches good device utilization only at a different batch size than the module preceding it, sequences yield after the upstream module so that the runtime can accumulate and \texttt{COMBINE} them into a batch downstream.
  For instance, attention saturates the GPU at a modest batch size while the sparsely activated MoE requires a much larger one.
  This applies to any sequence of modules with heterogeneous batch-size demand.
  Because the decisive factor is the model's compute characteristics,
  intra-forward yield points are selected once, statically, per model (\S\ref{subsec:static-plan}).
  \emph{Inter-forward} yield points occur across forward passes and are driven by runtime conditions rather than model structure. 
  A sequence yields either to achieve better utilization for batch progression (\eg switch from decode to prefill to refill sequence) or under resource pressure (\eg evict sequence under growing decode length that exceeds GPU memory).
  The yielded sequence can be combined again later with other active sequences running on GPU.
  Because these conditions depend on runtime state, inter-forward yield points are decided dynamically (\S\ref{subsec:continuous-batching}).
}

\mypar{Key development question: the choice of yield points} 
A central question arises when adopting sequence coroutines: 
given a model with multiple modules-and thus many candidate yield points-where should developers place yield points to maximize throughput?

\begin{figure}[t]
    \centering
    \includegraphics[width=\linewidth]{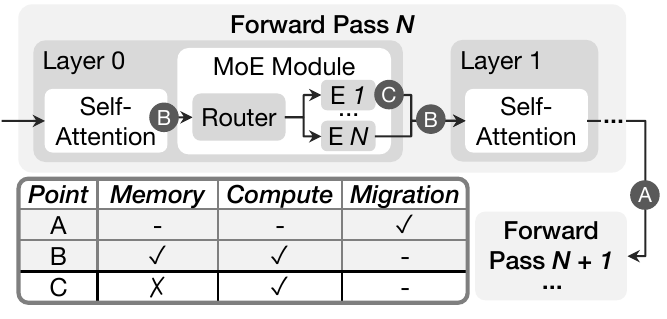}
    \caption{Yield point selection.}
    \label{fig:yield-point}
\end{figure}

For MoE architectures, we observe three practical and reusable yield-point options (shown in~\autoref{fig:yield-point}):
\myc{A}~treating the attention + MoE layers as a single coroutine unit,
\myc{B}~splitting the attention and MoE layers into two coroutine units, and
\myc{C}~treating each expert in an MoE layer as an independent coroutine unit.

These options reflect throughput-memory trade-offs:
\myc{A}~a combined attention-MoE coroutine minimizes intermediate states for sequence computation but limits expert-level batching and sequence concurrency;
\myc{B}~a split design allows the system to dynamically increase expert-level batch sizes without overwhelming memory usage; and
\myc{C}~a per-expert coroutine maximizes sequence computation concurrency but has prohibitive memory costs when initiating millions of coroutines.

According to these insights, we design the coroutine schedule to be hierarchical. For MoE batching, we adopt option~\myc{B}, inflating the batch size at MoE layers to achieve substantial throughput gains while keeping memory usage controlled at attention layers. 
To mitigate long-tailed decoding, we additionally employ option~\myc{A}. This enables lightweight migration and allows parallelism configurations to be adjusted dynamically through state checkpointing already made during yield.

The coroutine callbacks allow users to inject custom logic to handle sequence local state without modifying the runtime. These callbacks can be used to implement techniques such as dynamic quantization of KV-cache on a per-sequence basis~\cite{bitdecoding}, which \review{can} respond to runtime memory pressure or accuracy requirements. 

\section{System Design and Implementation}
\label{sec:batchgen-design}
\subsection{Sequence Coroutine Scheduling Algorithm}
\label{subsec:runtime-scheduler}

We design a sequence coroutine scheduler to sustain high compute efficiency across prefill and decoding. The core challenges stem from sparsely activated MoE layers, which require large batches, and from unpredictable long-tailed decoding. Our scheduler uses coroutine primitives to accumulate batches, adapt memory usage, and reduce long-tail inefficiency through continuous batching and dynamic parallelism.

\mypar{Providing sufficient batch size for sparse module}
Our scheduling algorithm aims to provide a sufficient batch size for MoE model by \texttt{COMBINE}-ing yielded sequences from attention.

The design is motivated by a key insight: \textit{a module-level roofline model enables static planning for batch accumulation}.  
Given the system's capacity for both active (GPU-resident) and inactive (host-resident) sequences, we can select the largest feasible module batch size that remains within memory constraints.

We design the coroutine scheduling in alignment with ~\autoref{fig:module-wrapper} for attention, MoE, and model forward pass as in ~\autoref{alg:batchgen-forward}, where the shaded area represents the code inside module wrapper that creates a coroutine.
For attention, the batch is split into sub-batches of size $B_{\text{attn}}$; each sub-batch executes attention, buffers its hidden states, optionally offloads KV-cache, and then yields.
For MoE, all sub-batch hidden states are combined into a large batch of size $B_{\text{moe}}$.

\begin{algorithm}[t]
\caption{Scheduling model forward pass}
\label{alg:batchgen-forward}
\small
\SetKwProg{Fn}{Function}{:}{}
\SetKwFor{ForEach}{foreach}{do}{end}
\SetKw{Yield}{\texttt{YIELD}}
\SetKw{Combine}{\texttt{COMBINE}}
\SetKw{Ret}{\texttt{return}}

\KwIn{
  $B_{\text{attn}}$: attention batch size (\texttt{COMBINE} on attention);\\
  $B_{\text{moe}}$: MoE batch size (\texttt{COMBINE} on MoE);\\
  Sequence group $\mathcal{S}$, $|\mathcal{S}|=B_{\text{moe}}$;
}
\KwOut{
  Final hidden states for all sequences in $\mathcal{S}$.
}

\Fn{\textsc{ForwardPass}$(\mathcal{S}, B_{\text{attn}}, B_{\text{moe}})$}{
    
    \ForEach{layer $\ell = 1$ \KwTo $L$}{
        Partition $\mathcal{S}$ into attention sub-batches of size $B_{\mathrm{attn}}$\;
        
        \ForEach{sub-batch $g \subseteq \mathcal{S}$}{
            \tikzmark{attn-start}
            $h_g \gets \textsc{Attention}(g, \ell)$ \tcp{Kept in GPU}
            \textsc{AsyncCkpt}($g$, $\ell$)\;
            \Yield $g$, $g.h \leftarrow h_g$\;
            \rlap{\tikzmark{attn-end}}

        }
        
        \tikzmark{start}
        $h_{\mathrm{combined}} \gets$ \Combine $\forall g.h, g\in\mathcal{S}$\;
        $h_{\mathrm{out}} \gets \textsc{MoE}(h_{\mathrm{combined}}, \ell)$ \tcp{Kept in GPU}
        \Yield $\mathcal{S}$, $\mathcal{S}.h \leftarrow h_{\mathrm{out}}$\; 
        \rlap{\tikzmark{end}}%

    }
    \Yield $\mathcal{S}$\;
}
\begin{tikzpicture}[remember picture,overlay]
    \fill[gray!15, opacity=0.4] 
        ([xshift=-2pt,yshift=6pt]pic cs:start) 
        rectangle 
        ([xshift=.7\linewidth,yshift=5pt]pic cs:end);
    \node[anchor=north east,font=\scriptsize] 
        at ([xshift=.7\linewidth,yshift=-14pt]pic cs:start) {MoE Coroutine};
\end{tikzpicture}
\begin{tikzpicture}[remember picture,overlay]
    \fill[gray!15, opacity=0.4] 
        ([xshift=-2pt,yshift=8pt]pic cs:attn-start) 
        rectangle 
        ([xshift=.7\linewidth,yshift=5pt]pic cs:attn-end);
    \node[anchor=north east,font=\scriptsize] 
        at ([xshift=.7\linewidth,yshift=-14pt]pic cs:attn-start) {Attn Coroutine};
\end{tikzpicture}
\end{algorithm}

\mypar{Mitigating long-tailed and unbalanced decoding}
Long-tailed decoding can only be managed at runtime, as output lengths are unknown in advance. We address this challenge through two mechanisms: \emph{Dynamic sequence management} maintains high GPU utilization by continuously refilling the batch with prefilled sequences. When sequences complete or are evicted, the scheduler invokes \texttt{COMBINE} to merge waiting sequences into the active batch, preventing GPUs from idling. We detail this in~\autoref{subsec:continuous-batching}. \emph{Straggler acceleration and load balancing} handle the final phase of batch completion. When only a few long-running sequences remain, \texttt{PARTITION} redistributes their computation across idle GPUs---using tensor parallelism for single stragglers or data parallelism for multiple. Throughout execution, \texttt{MIGRATE} rebalances sequences across nodes to prevent skew from uneven completion rates.

We illustrate the scheduling loop in~\autoref{alg:batchgen-runtime}. Nodes begin with prefilled sequences stored in host memory (line 1). The main loop (lines 2--6) repeatedly selects batches, executes forward passes, and invokes dynamic sequence management. The \textsc{OnRefillNode} callback (lines 7--11) triggers prefill when GPUs become underutilized. The \textsc{OnLongTail} callback (lines 12--14) detects stragglers and applies \texttt{PARTITION}, which requires yielding active sequences to reconfigure GPU memory for the new parallelism strategy \review{with its impact demonstrated in~\autoref{subsec:rl-rollout}}.

We design two callbacks, \textsc{OnRefillNode} and \textsc{OnLongTail}, to manage long-tail behavior. 
\textsc{OnRefillNode} supplies additional inactive sequences to the decoding phase so that idle or underutilized GPUs remain fully occupied (lines 7-11). 
When long-tail sequences persist and no inactive sequences remain to refill the batch, the scheduler invokes \textsc{OnLongTail} (lines 12-14). This callback waits for all active sequences on the target GPUs to yield, ensuring their states are safely checkpointed before reconfiguring model parallelism and redistributing the remaining work.

\begin{algorithm}[t]
\caption{Scheduling Sequence Computation}
\label{alg:batchgen-runtime}
\small
\SetKwProg{Fn}{Function}{:}{}
\SetKwFor{ForEach}{foreach}{do}{end}
\SetKwFor{While}{while}{do}{end}
\SetKwIF{If}{ElseIf}{Else}{if}{then}{else if}{else}{end}
\SetKw{Yield}{\texttt{YIELD}}
\SetKw{Combine}{\texttt{COMBINE}}
\SetKw{Migrate}{\texttt{MIGRATE}}
\SetKw{Partition}{\texttt{PARTITION}}

\KwIn{
$B_{\text{attn}}$: attention batch size;\\
$B_{\text{moe}}$: MoE batch size;\\
Global sequence batch $\mathcal{S}_{\text{global}}$, Node sequence batch $\mathcal{S}_{\text{node}}$ \\
}

Distribute $\mathcal{S}_{\text{global}}$ evenly\;

\While{$\exists s \in \mathcal{S}_{\text{node}}: \neg\textsc{Finished}(s)$}{
    \tcc{Run on each node}
    $\mathcal{S} \gets \textsc{SelectBatch}(\mathcal{S}_{\text{node}}, B_{\text{moe}}, \text{status=inactive})$\;
    \Combine $\mathcal{S}$ as active sequences\;
    \textsc{ForwardPass}($\mathcal{S}, B_{\text{attn}}, B_{\text{moe}}$)\;
    \textsc{DynamicSequenceManagement}($\mathcal{S}$)\;

    \If{\textsc{OnRefillNode}(nodes)}{
        wait until all $s \in \mathcal{S}_{\text{node}}$ \Yield\;
        \tcc{Trigger prefill temporally}
        $\mathcal{S} \gets \textsc{SelectBatch}(\mathcal{S}_{\text{node}}, B_{\text{moe}}, \text{status=init})$\;
        \Combine $\mathcal{S}$ as active sequences\;
        \textsc{ForwardPass}($\mathcal{S}, B_{\text{attn}}, B_{\text{moe}}$)\;
        \tcc{decoding continues here}
    }

    \If{\textsc{OnLongTail}(nodes, gpus)}{
        wait until all $s \in \mathcal{S}_{\text{node}}$ \Yield on target GPUs\;
        \Partition remaining sequences to target GPUs\;
    }

}
\end{algorithm}

\subsection{System Memory Model}
\label{subsec:system-components}

We design the system memory layout to support coroutine scheduling with minimal overhead. 
The key challenges arise from the different memory demands of prefill and decoding, the sensitivity of batch combination size to peak memory usage, and the need to shift efficiently between phases while preserving accumulated state. 

Our memory layout addresses these issues by unifying host and device memory management, paged KV-cache management, and providing phase-agnostic GPU buffers that can be reconfigured without disrupting coroutine execution. 

\review{The residence of the KV cache and model weights, i.e., whether each is offloaded to host memory, is a throughput-driven decision resolved by the coroutine scheduling plan (\autoref{subsec:static-plan}), depending on whether the host-device transfer cost is outweighed by the throughput gain from larger-batch execution.}

\begin{figure}[t]
    \centering
    \includegraphics[width=\linewidth]{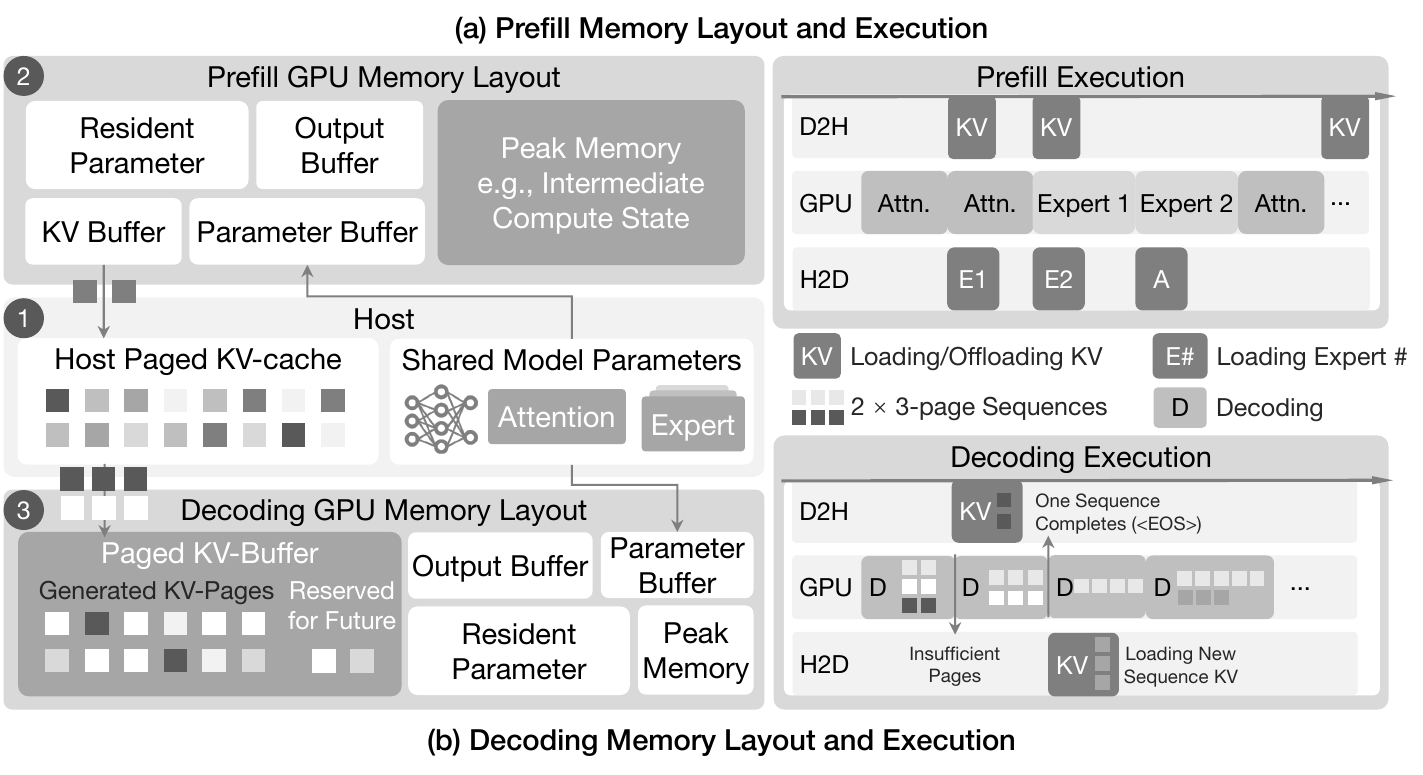}
    \caption{Memory layout and execution flow for prefill (top) and decode (bottom) phases. Host memory serves as checkpoint storage for parameters and KV-cache, while GPU memory is partitioned into persistent and transient regions.}
    \label{fig:memory-layout}
\end{figure}

\autoref{fig:memory-layout} illustrates \sys's memory layout and execution flow for both prefill and decode phases, rooted in the following principles.

\mypar{Host memory should serve as a unified store}
On each node, the host memory holds shared model parameters and KV-cache for all sequences scheduled to the node (in~\myc{1}).
This provides two benefits: (i)~\emph{a single source of truth} for all model states, avoiding redundant per-GPU copies, and (ii)~\emph{local coroutine store} for checkpointing KV-cache on \texttt{YIELD} and restoring it on \texttt{COMBINE} without distributed synchronization.
Parameter movement is unidirectional: host memory serves as the source for GPU prefetching, not a destination for offloading.

\mypar{Supporting local phase switching with fine-grained GPU buffers}
\sys partitions GPU parameter memory into two regions (in~\myc{2} and~\myc{3}).  
(i)~\textit{Resident parameters} store parameters that remain in the GPU throughout the current phase, which is typically small (\eg, LayerNorm).
During decoding, it may include all attention parameters when memory permits.
(ii)~\textit{Parameter and KV buffer} is a transient staging area for prefetching and offloading. 
The buffer slot ownership is released after the module completes execution.
As prefill and decoding share the same design, we reconfigure the size of these components during the phase swap.

\mypar{Increasing peak memory budget for prefill}
Prefill-phase attention exhibits high peak memory demand due to FlashAttention's $O(N)$ memory complexity, where $N$ is the sequence length. In contrast, decode-phase attention requires only $O(1)$ memory per sequence, as each iteration processes a single token against the existing KV-cache.
\review{We design the prefill phase to offload more aggressively by asynchronously offloading the microbatch KV-cache from each layer's attention to host and reclaiming its GPU buffer immediately. 
The prefill phase has enough computation to hide the offloading cost. At any instant the GPU therefore holds at most two layers' worth of KV-cache, making prefill memory less demanding from both batch size and sequence length.
This results in minimizing buffer size in prefill (in~\myc{2}).
}
The example for the ring parameter buffer and single KV buffer is shown in~\hyperref[fig:memory-layout]{\autoref*{fig:memory-layout}a}, with the running and prefetching expert using the ring buffer and KV-cache offloading (asynchronously after attention) using the single buffer.

\mypar{Increasing sequence concurrency for decoding}
Each GPU maintains a paged KV-cache manager (in~\myc{3}) that allocates and tracks KV pages at a fine granularity. 
 
During decoding, \sys employs a two-page buffer strategy: each sequence reserves only two KV pages for subsequent decoding iterations, with additional pages allocated on demand as the sequence grows. 
This lazy allocation maximizes the number of active sequences. 
\review{
Offloading during decoding considers a trade-off between parameters and KV-cache transfer cost and the throughput gain from larger batch size.
Decode has low arithmetic intensity thus offers little computation to hide host-to-device transfers.
We demonstrate a case for benefits under tight memory budgets (\S\ref{subsec:limited-memory}).
}

When GPU memory becomes insufficient, we evict sequences until two pages can be allocated for all active sequences.
Detailed design will be covered in~\autoref{subsec:continuous-batching}.

\subsection{Dynamic Sequence Management}
\label{subsec:continuous-batching}

The dynamic sequence management design is unique to batch inference: to maximize throughput, we must maintain large batch sizes or high parallelism throughout both prefill and decoding phases.

\sys checks memory state and adjusts the active batch every $P$ tokens (i.e., after decoding one KV page of tokens). At each page boundary, the scheduler executes four phases:
(i)~\textit{Sync}: wait for pending asynchronous KV append operations to complete---these operations continuously propagate newly generated KV entries from GPU to host memory, maintaining host as the single source of truth;
(ii)~\textit{Eviction}: \texttt{YIELD} completed sequences, releasing their GPU and host KV pages;
(iii)~\textit{Extension}: for sequences approaching their allocated page limit, either extend allocation if GPU pages are available, or \texttt{YIELD} them to host (marking as suspended) to free pages for other sequences;
(iv)~\textit{Refill}: \texttt{COMBINE} new sequences from the prefilled or suspended pool, launching async KV restoration that overlaps with the next page's forward passes.

These operations provide the following properties:

\mypar{Ensuring decoding batch size is sufficient}
\sys uses \textit{adaptive sequence selection}. When GPU memory is insufficient for all active sequences, \sys must decide which sequences to \texttt{YIELD}. 
The scheduler prioritizes yielding sequences with the most progress (highest decoded length), as they have more KV-cache already checkpointed to host and are closer to completion. This strategy maximizes the number of sequences that can make forward progress.

\mypar{Ensuring load-balance on refill}
Long-tail generation creates imbalance across GPUs as sequences complete at different times. 
When a GPU finishes sequences, \sys uses \texttt{MIGRATE} to redistribute suspended sequences from other nodes, keeping all GPUs utilized. Refill candidates are selected in FIFO order to preserve fairness. When all nodes lack sufficient suspended sequences, \sys temporarily switches to prefill mode to generate new sequences.
Note that \texttt{MIGRATE} requires the coroutine task queue to be blocked to (i) avoid races between active execution and memory transfer, and (ii) synchronize on the global count of inactive sequences. This blocking is low-cost since migration overlaps with sequence computation on other GPUs. 

\mypar{Ensuring single sequence straggler is parallelized}
When straggler sequences remain, batching benefits diminish and most GPUs become idle. 
\sys allows configuring a threshold on global batch size to trigger \texttt{PARTITION}: for multiple stragglers, sequences are distributed across GPUs using data parallelism; for a single straggler, computation is partitioned across GPUs using tensor parallelism. 
The KV-cache distribution depends on the attention architecture: for MLA, the compressed latent KV-cache is replicated across GPUs; for MHA/GQA, the KV-cache is split across attention heads.

\subsection{Optimization of Scheduling Plans}
\label{subsec:static-plan}

The runtime also follows a static plan that fixes batch sizes, buffer allocations, and yield points. 
Manual tuning is infeasible given the combinatorial space of model scales and hardware configurations, so \sys derives this plan automatically through lightweight profiling and simulation.

\mypar{Module-level performance modeling}
Modern LLMs are structurally homogeneous, allowing \sys to profile a single representative layer rather than the entire model. It measures attention, MoE, and collective communication kernels over multiple batch sizes and fits the results to roofline-style models that predict runtime as a function of batch size and memory allocation.

\mypar{Execution DAG construction}
\sys models a single layer's forward pass as a directed acyclic graph (DAG) to simulate execution under different configurations. Each node in the DAG represents either a computation (attention, MoE, \texttt{YIELD} checkpointing) or a data transfer (parameter prefetch, KV-cache offload/restore for \texttt{COMBINE}). Edges encode dependencies: a module cannot execute until its parameters are loaded; \texttt{COMBINE} cannot proceed until KV-cache restoration completes (if needed). Node costs are assigned from the profiled execution times and memory requirements at each candidate batch size. The critical path through this DAG---the longest dependency chain---determines the layer's execution time, capturing how computation and data movement overlap.

\mypar{Configuration search}
Given a DAG, finding the critical path requires a single topological traversal using dynamic programming, with complexity $O(V + E)$ where $V$ and $E$ are the number of nodes and edges. Since a layer's DAG contains fewer than 100 nodes, this computation is negligible. \sys enumerates candidate configurations ($B_{\text{attn}}$, $B_{\text{moe}}$, and buffer sizes), constructs the corresponding DAG for each, and selects the configuration with the shortest critical path.
\subsection{Coroutine Overhead Analysis}
\label{subsec:overhead}

A natural question is whether coroutine-based scheduling introduces noticeable overhead. 
Table~\ref{tab:overhead} summarizes the cost of all coroutine primitives. 

\begin{table}[t]
\centering
\small
\resizebox{\columnwidth}{!}{
\begin{tabular}{llr}
\toprule
\textbf{Category} & \textbf{Operation} & \textbf{Cost} \\
\midrule
\multirow{2}{*}{Intra-forward \texttt{YIELD}} 
  & Hidden state checkpoint & $<$5\,$\mu$s \\
  & KV-cache offload & overlapped \\
\midrule
\multirow{2}{*}{Inter-forward \texttt{YIELD}}
  & CUDA event sync & $<$1\,$\mu$s \\
  & Metadata update & $<$10\,$\mu$s \\
\midrule
\multirow{2}{*}{\texttt{COMBINE}} 
  & Without offloading & 0 (GPU resident) \\
  & With offloading & $\sim$0.2\,ms per seq. per layer\\
\midrule
\texttt{PARTITION} & Parallelism reconfig & $\sim$5--10\,s (when triggered) \\
\midrule
\texttt{MIGRATE} & KV-cache transfer & Opportunistic, overlapped \\
\midrule
\multirow{2}{*}{Scheduling}
  & Intra-forward & 0 (static plan) \\
  & Cross-node sync & 5--10\,ms / 64 tokens \\
\bottomrule
\end{tabular}
}
\caption{Coroutine primitive overhead for DeepSeek-R1 on 8$\times$H20 with 10K context length and batch size 512.}
\label{tab:overhead}
\end{table}

\mypar{Frequently invoked primitives have negligible overhead}
Coroutine boundaries occur once per module. Hidden-state checkpointing writes only $\sim$7\,MB for 512 sequences (7168-dim), completing in under 5\,$\mu$s at HBM bandwidth, compared to $\sim$3\,ms for the module's computation. KV-cache offloading is issued asynchronously and fully overlaps with MoE execution. Between forward passes, \texttt{YIELD} performs a CUDA event sync ($<$1\,$\mu$s) and updates sequence metadata on the CPU ($<$10\,$\mu$s for 512 sequences).

\texttt{COMBINE} overhead depends on whether KV-cache offloading is active. Without offloading, combined sequences are already GPU-resident, incurring zero cost. With offloading, \texttt{COMBINE} must restore KV-cache from host memory; for a 10K-token sequence this takes $\sim$0.2\,ms per layer over PCIe\,5.0. This cost arises only when refilling batches and is amortized by the larger batches that offloading enables.

\mypar{Primitives with millisecond-level overhead are negligible to overall computation time}
\review{\texttt{MIGRATE} is invoked opportunistically to balance the number of inactive sequences across nodes. 
The number of migrations over a batch stays small relative to the total sequence count and the migration cost can be smaller than prefill.}
When \texttt{MIGRATE} is scheduled in advance, the KV-cache transfer is fully overlapped with ongoing computation on other sequences, adding zero latency to the critical path. 

  \review{For instance, on a standard H100 HGX node with 400\,Gb/s ($\sim$50\,GB/s) InfiniBand link, prefilling a 2K-token DeepSeek-R1 sequence costs $\sim$112\,ms even at the GPU's peak throughput (BF16 MLA attention at 989\,TFLOP/s plus FP8 MoE GEMMs at 1979\,TFLOP/s), whereas migrating the sequence's entire 144\,MB MLA KV-cache over that link takes only $\sim$2.9\,ms, which is less than 3\% of the prefill compute it overlaps with.}
Although \texttt{PARTITION} takes seconds, it is triggered only when long-tail sequences are detected (typically once per batch near completion) and enables idle GPUs to accelerate remaining stragglers. Meanwhile, batch completion time is more than 10 minutes.
Dynamic scheduling occurs only at page boundaries (every 64 tokens), requiring 5--10\,ms for cross-node synchronization---approximately 0.1--0.2\% of the compute time for 64 decode iterations.

\subsection{Implementation Details}
\label{subsec:implementation}

\begin{figure}[!t]
    \centering
    \includegraphics[width=\linewidth]{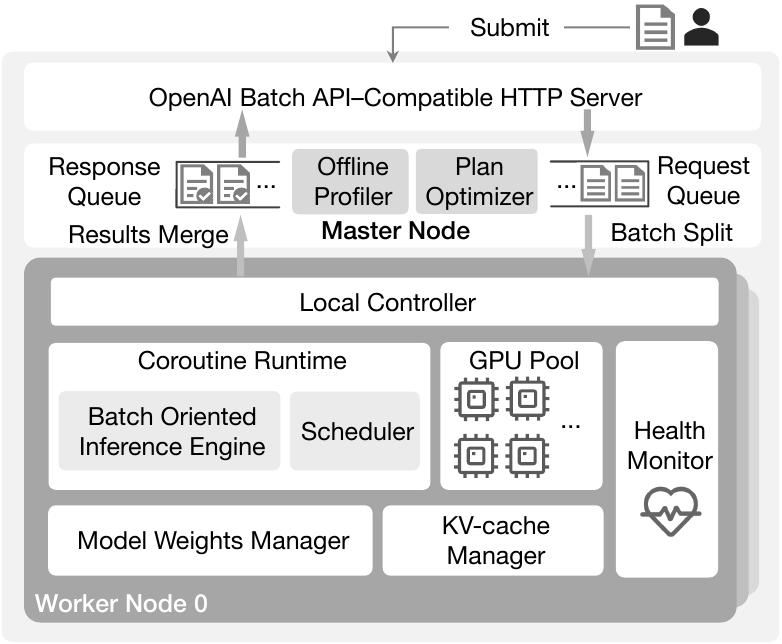}
    \caption{\sys system architecture with master-worker organization, API endpoints, and inference engine.}
    \label{fig:implementation}
\end{figure}

We implement \sys in 13K LoC of C++ and 49K LoC of Python. \review{The C++ layer implements the host-side KV-cache system and the asynchronous host--device copy engine, while the Python layer implements the coroutine scheduler and the PyTorch-based forward-pass skeleton for each model. Beyond third-party kernels (e.g., FlashAttention~\cite{dao2022flashattention}, DeepGEMM~\cite{deepgemm}), 
\sys develops and maintains its own library of kernels for batch inference, adapted to \sys's KV-cache and input/output buffer layout.

}
\autoref{fig:implementation} shows the \sys system architecture for deployment at scale \review{with the control plane implemented on an industry-customized Ray~\cite{ray}}. Key features include:

\mypar{OpenAI-compatible batch API}
\sys exposes an HTTP endpoint compatible with the OpenAI Batch Inference API~\cite{openaibatch}. A master node per model-parallel group accepts batch submissions, partitions sequences across workers via the coroutine scheduler, and returns results preserving input order. This enables \sys to serve as a drop-in replacement for existing batch inference pipelines.

\mypar{Batch-oriented inference engine}
The inference engine integrates high-performance compute and communication kernels (e.g., DeepEP~\cite{deepep}) and includes additional kernel fusions tailored for model types. It supports phase-specific parallelism strategies:  
(i)~\textit{DP prefill with offloading}, where prefill is compute-bound and parameter/KV transfers are fully hidden, enabling scalable data parallelism; and  
(ii)~\textit{DP+EP decoding}, which combines data-parallel attention with expert-parallel MoE to maximize bandwidth utilization during generation.

\mypar{Cold-start optimization}
Batch inference often runs on spot instances where cold-start latency impacts usable compute time. \sys includes a \textit{model weights manager} with:
(i)~\textit{Huge pages}: Allocating GPU-accessible host memory incurs page faults proportional to pool size. \sys uses 2\,MB huge pages, reducing initialization time from minutes to seconds for TB-scale pools.
(ii)~\textit{Fast checkpoint loading}: \sys adopts the memory-mapped checkpoint format from ServerlessLLM~\cite{serverlessllm}, enabling rapid weight loading that fits within typical spot instance lifetimes.

\mypar{Failure detection and recovery}
At scale, GPU and node failures are inevitable. \sys implements \textit{health monitoring} where per-node responses carry all GPU statuses. For recovery, \texttt{MIGRATE} can either transfer KV-cache state or trigger recomputation. Since migrating hundreds of gigabytes may be slower than regenerating, \sys uses its performance model to estimate both costs and selects the faster path.
\section{Evaluation}
\mypar{Testbed}
We evaluate \sys on a production-grade multi-GPU cluster representative of real LLM deployment environments. The testbed includes NVIDIA H20 and H200 servers scaled from 8 to 128 GPUs. Each node offers high-bandwidth intra-node connectivity via NVLink and PCIe\,5.0, with 2\,TB of host memory, and nodes are interconnected through a 200\,Gb/s RDMA-enabled InfiniBand fabric, matching common datacenter configurations for large-scale model serving. The H200 servers are provided by the University of Edinburgh.

\mypar{Baselines}
Our primary baselines are SGLang (v0.5.5.post3)~\cite{sglang} and vLLM (v0.11.2)~\cite{vllm}, which are widely recognized as SOTA open-source systems.
We evaluate SGLang-Optimized, which applies exhaustive tuning for maximum throughput: we configure 16 DP-attention ranks (a throughput-friendly parallelism strategy), tune the memory allocation fraction to the threshold before OOM, and selectively restrict CUDA graph capture to necessary batch sizes to reduce GPU memory consumption and allow larger runtime batches.

For memory-constrained scenarios (\autoref{subsec:limited-memory}), we compare against systems with offloading optimizations: DeepSpeed~\cite{deepspeed}, FlexGen$^\dagger$~\cite{flexgen}, and MoE-Lightning$^\dagger$~\cite{moe-lightning}.
For disaggregated inference (\autoref{subsec:real-world}), we evaluate prefill-decode (PD) disaggregation using SGLang's implementation. We exclude attention-expert disaggregation~\cite{megascale-infer} due to the lack of open-source implementations.

\mypar{Models}
Our evaluation spans four SOTA MoE models, including Mixtral-8$\times$7B (47B parameters) and Mixtral-8$\times$22B (141B parameters)~\cite{jiang2024mixtralexperts},
DeepSeek-R1 (671B parameters)~\cite{deepseek-r1}, and
Kimi-K2 (1T parameters)~\cite{kimi-k2}.

\subsection{Application 1: Offline Inference}
\label{subsec:offline-inference}

We evaluate \sys using LongBench~\cite{longbench}.
We evaluate on two common production patterns: 
(i)~\emph{prefill-heavy} (8K input, 2K output tokens): reflects summarization and information extraction tasks,
and (ii)~\emph{decoding-heavy} (2K input, 8K output tokens): captures reasoning-intensive workloads such as chain-of-thought question answering.
~\autoref{tab:offline-inference} reports batch completion time for 6K sequences.

\begin{table}[t]
    \centering
    \resizebox{\columnwidth}{!}{
    \begin{tabular}{ll cc cc cc}
    \toprule
    & & \multicolumn{2}{c}{\textbf{H20 (8 GPUs)}} & \multicolumn{2}{c}{\textbf{H20 (16 GPUs)}} & \multicolumn{2}{c}{\textbf{H200 (8 GPUs)}} \\
    \cmidrule(lr){3-4} \cmidrule(lr){5-6} \cmidrule(lr){7-8}
    \textbf{Model} & \textbf{Framework} & 8K-2K & 2K-8K & 8K-2K & 2K-8K & 8K-2K & 2K-8K \\
    \midrule
    \multirow{5}{*}{\parbox{2.2cm}{\centering DeepSeek-R1\\(671B)}}
        & vLLM              & 1020.6   & 3233.8    & 373.8 & 746.3  & 1285.2 & 1076.7 \\
        & SGLang            & 816.7 & 3116.7 & 375.0 & 981.9  & 258.9  & 880.8 \\
        & SGLang-Opt        & \underline{816.7} & \underline{3116.7} & \underline{137.0} & \underline{242.5}  & \underline{95.2}   & \underline{208.3} \\
        & \textbf{\sys}     & \textbf{625.1} & \textbf{1684.1} & \textbf{82.6} & \textbf{194.3} & \textbf{75.1} & \textbf{165.0} \\
        & \textit{Speedup}  & \textit{1.31$\times$} & \textit{1.85$\times$} & \textit{1.66$\times$} & \textit{1.25$\times$} & \textit{1.27$\times$} & \textit{1.26$\times$} \\
    \midrule
    \multirow{5}{*}{\parbox{2.2cm}{\centering Kimi-K2\\(1T)}}
        & vLLM              & OOM   & OOM    & 1862.0 & 2159.2 & 3474.5 & 3480.5 \\
        & SGLang            & OOM   & OOM    & 447.9  & 1468.1 & 293.8  & 975.6 \\
        & SGLang-Opt        & OOM   & OOM    & \underline{150.6}  & \underline{430.6}  & \underline{167.5}  & \underline{525.0} \\
        & \textbf{\sys}     & \textbf{659.8} & \textbf{1693.6} & \textbf{116.8} & \textbf{317.5} & \textbf{125.0} & \textbf{392.0} \\
        & \textit{Speedup}  & --    & --     & \textit{1.29$\times$} & \textit{1.36$\times$} & \textit{1.34$\times$} & \textit{1.34$\times$} \\
    \midrule
    \end{tabular}
    }
    \caption{Batch completion time (minutes) for 6K sequences on LongBench. OOM indicates out-of-memory. Speedup is relative to SGLang-Optimized. For Kimi-K2 on 8$\times$H20, \sys is the only system that completes execution.}
    \label{tab:offline-inference}
\end{table}

\mypar{\sys achieves larger speedup on weaker GPUs}
On 16$\times$H20 and 8$\times$H200 clusters, \sys delivers 1.25--1.66$\times$ speedup over SGLang-Optimized across all configurations. 
The improvement is consistently larger on H20 than on H200. 
H200's higher compute throughput shrinks the relative cost of prefill, where \sys's batch accumulation provides the most benefits. On H20, prefill dominates end-to-end time, and \sys's ability to form large expert batches yields proportionally greater gains.

\mypar{\sys enables larger models and far larger batches under tight memory budgets}
On 8$\times$H20 GPUs (768\,GB total HBM), baseline behavior diverges by model. DeepSeek-R1's parameters (approximately 642\,GB in FP8) fit in HBM, but baselines must partition remaining capacity between KV-cache and working memory, limiting batch sizes to 8--16 sequences. SGLang-Optimized cannot improve over default SGLang in this regime---exhaustive tuning yields no additional headroom when memory is saturated. \sys achieves 1.31--1.85$\times$ speedup by offloading KV-cache to host memory, enabling batch sizes of 1800+ sequences. For Kimi-K2 (1T parameters), no baseline completes execution: the model exceeds HBM capacity entirely. \sys is the only system that runs Kimi-K2 on this hardware configuration.

The offloading regime exhibits a counterintuitive effect: \sys delivers nearly identical throughput on DeepSeek-R1 and Kimi-K2, even though Kimi-K2 is about 1.5$\times$ larger. With offloading, performance becomes PCIe-bandwidth-bound rather than compute-bound. Because both models have similar per-token KV-cache sizes and PCIe bandwidth is fixed, their achievable batch sizes-and thus throughput-converge. This trade-off remains advantageous: offloading enables much larger expert batches, keeping MoE layers in the memory-bound regime where compute scales sublinearly with batch size, while greater sequence-level parallelism amortizes transfer costs.

\mypar{Comparison with TensorRT-LLM}
We additionally compare against TensorRT-LLM~\cite{trt-llm}, which represents state-of-the-art kernel optimization but requires non-trivial deployment effort. On 16$\times$H20 GPUs with the 8K-2K workload, \sys outperforms TensorRT-LLM by 10\%, demonstrating that coroutine-based scheduling provides benefits orthogonal to kernel-level optimizations.

\subsection{Application 2: Test-Time Scaling}
\label{subsec:test-time-scaling}

Unlike offline batch inference, test-time scaling places SLO constraints on batch completion time to maintain acceptable user experience, requiring systems to choose batch sizes that finish before the deadline. We use Recursive Self-Aggregation (RSA)~\cite{rsa} as a representative workload. After the first round, each subsequent round combines $K$ prior answers as the prefill input, producing a prefill-to-decode ratio of $K$:1. We evaluate this setting on DeepSeek-R1 using 16$\times$H20 GPUs with a fixed 8K decoding length per round.

\mypar{\sys enables efficient SLO--throughput tradeoffs}
~\autoref{tab:test-time-scaling} reports the number of sequences served under 30-minute and 60-minute SLO targets. \sys processes 1.25--1.57$\times$ more sequences under the 30-minute SLO and 1.66--1.75$\times$ more under the 60-minute SLO compared to SGLang-Optimized. The larger gains under relaxed SLOs reflect \sys's ability to form larger batches when deadlines permit, improving expert-level compute density. Prefill-heavy configurations (higher $K$) show greater improvements because compute-bound prefill enables effective overlap of parameter and KV-cache transfers at no additional cost.

\mypar{\sys maintains advantages even under constrained batch sizes}
Under tighter SLOs, \sys cannot always operate at its theoretically optimal batch configuration. For instance, in the (T=4, N=8, K=4) 30-minute scenario, meeting the completion deadline requires reducing the MoE batch size below the compute-saturating threshold. This sub-optimal configuration explains why \sys's speedup at 30 minutes (1.25$\times$) is lower than at 60 minutes (1.66$\times$)---the SLO constraint limits batching flexibility. Nevertheless, \sys still outperforms all baselines, demonstrating that its coroutine-based scheduling provides benefits even when batch sizes are externally constrained. This makes \sys suitable for online services with relaxed (minute-to-hour scale) SLOs where some latency--throughput tradeoff is acceptable.

\begin{table}[t]
    \centering
    \resizebox{\columnwidth}{!}{
    \begin{tabular}{ll cc cc cc}
    \toprule
    & & \multicolumn{2}{c}{\textbf{T=4, N=8, K=4}} 
      & \multicolumn{2}{c}{\textbf{T=2, N=16, K=4}} 
      & \multicolumn{2}{c}{\textbf{T=3, N=8, K=2}} \\
    \cmidrule(lr){3-4} \cmidrule(lr){5-6} \cmidrule(lr){7-8}
    \textbf{Model} & \textbf{Framework} 
        & 30min & 60min 
        & 30min & 60min 
        & 30min & 60min \\
    \midrule
    \multirow{4}{*}{\parbox{2.0cm}{\centering DeepSeek-R1}}
    & vLLM             & 10.9 & 23.1 & 10.7 & 20.1 & 7.3  & 15.0 \\
    & \review{SGLang-Opt}           & \underline{12.8} & \underline{25.9} 
    & \underline{13.4} & \underline{24.7} 
    & \underline{34.0} & \underline{69.2} \\
        & \textbf{\sys}    & \textbf{16} & \textbf{43} 
   & \textbf{21} & \textbf{42} 
   & \textbf{53} & \textbf{121} \\
    & \textit{Speedup} & \textit{1.25$\times$} & \textit{1.66$\times$} 
   & \textit{1.57$\times$} & \textit{1.70$\times$} 
   & \textit{1.56$\times$} & \textit{1.75$\times$} \\
    \midrule

    \end{tabular}
    }
\caption{Number of sequences served under SLO constraints for RSA test-time scaling on DeepSeek-R1 with 16$\times$H20 GPUs. RSA uses $T$ rounds, $N$ candidates per round, selects $K$ candidates for the next round, and fixes decoding length at 8K. Higher is better. Speedup is relative to SGLang-Optimized.}
\label{tab:test-time-scaling}
\end{table}

\subsection{Application 3: RL Training Acceleration}
\label{subsec:rl-rollout}

Reinforcement learning from human feedback (RLHF)~\cite{rlhf-rollout} is a standard stage in LLM training, and its rollout phase often dominates total training time. Unlike offline inference, RL rollout fixes the batch size (typically 256--1024)~\cite{hu2025openrlhf}, removing the primary lever used in batch inference: enlarging batches to improve expert utilization.

A major bottleneck in rollout is straggler sequences. Long-tailed generation lengths create synchronization barriers in synchronous training, forcing all workers to wait for the slowest trajectories. As the batch drains, GPU utilization collapses: once only a few sequences remain, most devices idle.

\sys mitigates stragglers via its coroutine callbacks. The \textsc{OnLongTail} callback triggers when the number of active sequences falls below a threshold, allowing the runtime to apply \texttt{PARTITION} and redistribute remaining work across idle GPUs, switching from data to tensor parallelism for single long-tail sequences. The same mechanism supports alternatives such as FP8 decoding or speculative decoding without changing the core runtime.

We evaluate on VeRL~\cite{verl} with DeepSeek-R1 on 16$\times$H20 GPUs. In a typical setting, each RL iteration uses 256 prompts with one response
  per prompt, for 256 rollout sequences in total, evenly distributed across the 16 GPUs at 16 active sequences per GPU initially. Generation lengths
  are heavily long-tailed, so the batch drains unevenly: \textsc{OnLongTail} triggers once the global active batch falls to $\leq 8$ sequences whose
  generation length exceeds 40K tokens. In our traces these final sequences account for 30--80\% of rollout time, yet leave at most 0.5 active
  sequences per GPU on average, exposing substantial idle capacity. \sys reclaims this capacity by applying \texttt{PARTITION} with FP8 decoding to
  the stragglers, reducing per-iteration rollout time by 5--10\%, with the gain increasing as the final generation length grows, consistent
  with~\cite{rollpacker}. Since rollout accounts for 60--80\% of training time, these savings translate directly into shorter end-to-end training.

\subsection{Scaling to Large Deployments}
\label{subsec:real-world}

We evaluate \sys at production scale with up to 128 GPUs using two trace-derived workloads: a prefill-heavy setting (12K input, 4K output tokens) and a balanced setting (6.5K input, 2.8K output). To disentangle runtime gains from kernel improvements, we also report results for \sys*, which replaces \sys's kernels with \review{SGLang's} while preserving the coroutine scheduler.

\autoref{tab:large-scale} shows end-to-end time for 10K requests.
Because the evaluated SGLang and vLLM versions exhibit stability issues when scaling DeepSeek-R1 beyond two nodes (16 GPUs), we extend them to 32-128 GPUs by running multiple independent 16-GPU data-parallel groups and aggregating throughput. \sys achieves 1.71-1.82$\times$ speedup on the 12K-4K workload and 2.2-2.3$\times$ on the 6.5K-2.8K workload relative to SGLang-Optimized.

\begin{table}[t]
    \centering
    \setlength{\tabcolsep}{8pt}
    \small
    \begin{tabular}{c l ccc}
    \toprule
    & & \multicolumn{3}{c}{\textbf{H20 GPUs}} \\
    \cmidrule(lr){3-5}
    \textbf{Workload} & \textbf{Framework} & 32 & 64 & 128 \\
    \midrule
    \multirow{4}{*}{12K-4K}
        & vLLM         & 866.0 & 437.2 & 218.9 \\
        & \review{SGLang-Opt}       & 239.2 & 120.8 & 61.1 \\
        & \textbf{\sys} & \textbf{139.7} & \textbf{66.3} & \textbf{33.5} \\
        & \textbf{\sys$^*$} & \textbf{110.7} & \textbf{52.6} & \textbf{26.3} \\
    \midrule
    \multirow{4}{*}{6.5K-2.8K}
        & vLLM         & 236.9 & 120.5 & 61.1 \\
        & \review{SGLang-Opt}       & 139.9 & 71.2 & 36.0 \\
        & \textbf{\sys} & \textbf{63.9} & \textbf{30.4} & \textbf{15.5} \\
        & \textbf{\sys$^*$} & \textbf{59.4} & \textbf{28.3} & \textbf{14.6} \\
    \bottomrule
    \end{tabular}
\caption{End-to-end time (minutes) for processing 10K requests on DeepSeek-R1. \sys$^*$ uses SGLang kernels with the \sys coroutine runtime.}
\label{tab:large-scale}
\end{table}

The performance gains from scaling stem from increased batch sizes: more GPUs provide additional memory capacity to accumulate larger batches before MoE execution, improving expert utilization. However, scaling exhibits sub-linear speedup due to communication overhead in MoE layers---the all-to-all collective for expert routing grows with GPU count. In our implementation, gains plateau beyond 64 GPUs; for the 128-GPU configuration, we deploy two 64-GPU instances and partition the input batch across them.

\mypar{Scaling beyond single-instance limits}
For a fixed model and parallelism strategy, there is a GPU count beyond which adding devices reduces per-GPU efficiency, as MoE all-to-all communication grows while computation per GPU stays constant. \sys treats this saturation point as the natural size of a single instance. Larger deployments are formed by replicating multiple independent instances, with an upper-level scheduler partitioning inputs and aggregating outputs. This hierarchical design preserves per-instance efficiency while scaling throughput linearly with instance count.

We compare against prefill-decode (PD) disaggregation~\cite{distserve} on SGLang at 128 GPUs. Each P or D unit consists of 16 GPUs, yielding seven P:D ratios.~\autoref{tab:scalability-pd} shows that PD disaggregation performance varies by 3.6$\times$ across configurations (38.6 to 137.9 minutes), requiring exhaustive profiling to identify the optimal ratio. \sys outperforms even the best PD configuration (4:4) by 2.2$\times$ without manual tuning.

\mypar{Production deployment}
  \sys is deployed as the batch-inference engine on an industry partner's production cluster. It runs as independent \sys instances launched and managed by an industry-customized Ray~\cite{ray}
  orchestrator. Each instance exposes an OpenAI-compatible API, and the scheduler intentionally over-subscribes it, dispatching far more requests than its concurrent capacity.
  Over-subscription serves two purposes. First, it keeps a large pool of sequences resident on each instance, giving the \sys runtime freedom to form batches by \texttt{COMBINE}-ing
  sequences selected from the pool rather than in arrival order (\S\ref{subsec:seq-coroutine-mechanisms}), which maximizes expert-level batching. Second, it keeps the instance
  continuously saturated: as sequences complete, the runtime immediately backfills from the resident pool, so concurrency stays high and the GPUs never drain. Completed requests are
  streamed to an instance-local directory that the scheduler polls periodically to collect results and dispatch new work.

\begin{table}[t]
    \centering
    \small
    \resizebox{\columnwidth}{!}{
    \begin{tabular}{l ccccccc}
    \toprule
    {P:D Ratio} & 1:7 & 2:6 & 3:5 & 4:4 & 5:3 & 6:2 & 7:1 \\
    \midrule
    PD Disagg. (min) & 137.9 & 67.4 & 49.0 & 38.6 & 40.1 & 55.6 & 112.0 \\
    \midrule
    \textbf{\sys} (min) & \multicolumn{7}{c}{\textbf{17.5}} \\
    \midrule
    \textit{Speedup} & \textit{7.9$\times$} & \textit{3.9$\times$} & \textit{2.8$\times$} & \textit{2.2$\times$} & \textit{2.3$\times$} & \textit{3.2$\times$} & \textit{6.4$\times$} \\
    \bottomrule
    \end{tabular}
    }
\caption{Comparison with PD disaggregation on 128$\times$H20 GPUs (10K requests, 8K-2K workload). Each ratio unit is 16 GPUs. \sys requires no ratio tuning.}
\label{tab:scalability-pd}
\end{table}

\subsection{Working with Limited Memory}
\label{subsec:limited-memory}

We evaluate \sys's offloading capability on commodity hardware by running large-scale MoE inference on a single NVIDIA A5000 GPU (24\,GB) with a 28-core AMD 7453 CPU and 1\,TB of host memory. This setup highlights \sys's ability to execute models far beyond GPU capacity.

As shown in \autoref{tab:limited-memory}, we report end-to-end processing time on GSM8K (512$\rightarrow$256) and ChatBotArena (256$\rightarrow$512). \review{FlexGen only supports dense models and} MoE-Lightning is not open-sourced, so we reproduce their offloading strategies based on the original papers and integrate them into our runtime$^\dagger$. For the single-GPU case, \sys additionally offloads attention computation to the CPU to relieve GPU memory pressure; this optimization is disabled in multi-GPU settings where CPU bandwidth is insignificant relative to total GPU resources$^\ddagger$.

\begin{table}[t]
    \centering
    \resizebox{\columnwidth}{!}{
    \begin{tabular}{rcccccc}
    \toprule
    $\rightarrow$Models &\multicolumn{2}{c}{Mixtral-8$\times$7B} & \multicolumn{2}{c}{Mixtral-8$\times$22B} & \multicolumn{2}{c}{DeepSeek-R1-671B}   \\
    \cmidrule(lr){2-3} \cmidrule(lr){4-5} \cmidrule(lr){6-7}
    $\rightarrow$Datasets   &GSM8K   & ChatBot &GSM8K   & ChatBot &GSM8K   & ChatBot\\
    \midrule
    vLLM            &20.4h           &N/A         &375.0h     &N/A        &N/A    &N/A   \\
    DeepSpeed       &26.0h           &187.4h          &116.0h     &950.4h     &N/A &N/A   \\
    FlexGen$^\dagger$         &18.8h           &156.0h          &83.5h &789.0h                 &N/A &N/A  \\
    MoE-Lightning$^\dagger$   &\underline{7.3h}            &\underline{58.5h}           &\underline{32.5h} &\underline{295.0h}                     &N/A &N/A     \\ 
    \midrule
    \textbf{\sys}$^\ddagger$   &\textbf{1.7h}  &\textbf{10.0h} &\textbf{5.1h}&\textbf{30.8h}  &\textbf{41.3h} &\textbf{328.5h}      \\
    \bottomrule
    \end{tabular}
    }
    \caption{Time to complete GSM8K (8.5K samples) and ChatBotArena (36K samples) on a single A5000 GPU (24\,GB). N/A indicates the system either runs out of memory or would take longer than 1000 hours to complete. Best results in \textbf{bold}, second best \underline{underlined}.}
    \label{tab:limited-memory}
\end{table}

\mypar{\sys achieves significantly higher efficiency under limited resources}
Under constrained hardware settings where most baselines cannot complete the dataset within a reasonable time, \sys achieves up to a $9.6\times$ speedup. As model size increases, baseline systems fail entirely, while \sys continues to finish the workload. 
FlexGen and MoE-Lightning suffer from extremely small per-expert batch sizes--40-1000$\times$ smaller than what is required to fully utilize the GPU--because they retain full MoE layers on device, capping the memory available for batching. In contrast, \sys accumulates large batches via coroutine \emph{yields} and fully offloads experts, enabling per-expert batch sizes that reach the compute-bound regime. vLLM remains bottlenecked by on-demand parameter fetching, whereas \sys closes this gap by fully overlapping parameter transfers with computation in the coroutine runtime and its buffer.

\section{Limitations and Future Work}
    \mypar{Batch-size regime and best-effort operation}
  For a given model, its compute
  characteristics determine an optimal batch size at the \texttt{COMBINE} point that saturates the device. In practice, however, this saturation point
  cannot always be reached: even when enough candidate requests are available to batch, the system's memory capacity may be insufficient to hold
  them, in which case \sys forms the largest batch memory permits and operates below saturation on a best-effort basis. \autoref{fig:expert-batch}
  illustrates this regime: saturating the compute of every expert requires 16384 tokens to reach the MoE gate. This target is comparatively easy to
  meet during prefill, where a single sequence of length 16384 suffices, but during decoding it corresponds to a global batch of 16384 concurrent
  requests---rarely attainable given the long-context requests typical of production. Decoding therefore runs in the best-effort
  regime. In our limited-memory evaluation (\autoref{subsec:limited-memory}), a batch size of 6000 still yields gains, bounded by host memory.
  
    \mypar{Generalization and future work}
  The coroutine abstraction is general: it applies to any generation workload whose stages have distinct compute characteristics and memory requirements. Our current
  implementation targets transformer MoE models, a widely deployed architecture that benefits directly from intra-model yielding at attention--MoE boundaries. Extending \sys to other models and tasks is future work. Within a single model, vision-language models are a natural candidate: image
  encoding is substantially more compute-heavy than the language stages, so a yield point after the encoder could let the runtime rebatch the lighter
  stages. The abstraction also extends across models: where a generation pipeline connects multiple models with distinct computation
  characteristics---as in omni-modal systems that couple a language model with separate speech or vision models~\cite{xu2025qwen25omnitechnicalreport, yin2026vllmomnifullydisaggregatedserving}---each
  model can be treated as a coroutine task, applying yield and \texttt{COMBINE} at model boundaries to batch each at its own optimal granularity.

\section{Related Work}
\mypar{Throughput-optimized offloading systems}
Current offloading systems have not targeted sparse modules and long-tailed decoding.
FlexGen~\cite{flexgen} is designed for dense modules, thus treating MoE layer as a single dense layer for offloading. MoE-Infinity~\cite{xue2024moe} is designed for MoE inference on personal machines with a batch size of one.  DeepSpeed~\cite{deepspeed} and MoE-Lightning~\cite{moe-lightning} are agnostic to expert batching, with a global batch size on the whole model decided. PowerInfer~\cite{powerinfer} and Fiddler~\cite{fiddler} partition computation between the GPU and CPU, with lightweight attention and expert compute running on CPU, aiming for small batch sizes where compute is not intensive. MoE-CAP~\cite{jiang2026moe} is for benchmarking MoE systems under different scenarios.

\mypar{Continuous batching for high throughput}
Continuous batching (used in vLLM~\cite{vllm}, Orca~\cite{orca}, and Llama.cpp~\cite{ollama}) was introduced to address long-tail related TTFT concerns in interactive inference.
These frameworks insert small prefill batches into the decoding phase.
vLLM~\cite{vllm}, Orca~\cite{orca}, and Llama.cpp~\cite{ollama} follow this strategy directly.
NEO~\cite{neo} interleaves prefill and decoding across GPU and CPU resources, while systems such as BlendServe~\cite{blend-serve} and other micro-batching approaches~\cite{stream-batch} share the GPU in the temporal domain. 
The effective average batch size over the entire execution becomes even smaller, limiting throughput.

\mypar{Batching in training systems}
Training fixes batch sizes to ensure gradient stability and convergence, while inference can vary batch sizes freely~\cite{mai2019taming}.
In addition, the training systems are optimized for prefill and gradient updates~\cite{alpa}, and decoding speed is not a factor.
The MoE training line of work~\cite{janus, lina, 258904, smart-moe, wagenlander2024tenplex} is orthogonal to \sys, although \sys is compatible with the fixed batch size in RL post-training of LLMs~\cite{rlhf-rollout, pmlr-v202-wang23aj}.

\section{Conclusion}
 
This paper introduces the event-driven sequence coroutine architecture, a new computation model for large-scale batch inference. By enabling fine-grained yielding, combining, partitioning, and migration of sequence computation, \sys overcomes structural bottlenecks in sparse models and long-tail decoding, delivering substantial improvements in batch completion time across various applications. Beyond immediate throughput gains, our results suggest broader implications: extending coroutine abstractions to multimodal workflows, and co-designing the coroutine with the underlying high-performance GPU kernels. We believe this architecture opens a rich systems research agenda for the next generation of AI engine design.

\bibliographystyle{plain}
\bibliography{main}


\end{document}